\documentclass[twocolumn,showpacs,preprintnumbers,amsmath,amssymb]{revtex4}

\usepackage{graphicx}

\usepackage{dcolumn}
\usepackage{bm}

\newcommand{\be}{\begin{equation}}
\newcommand{\ee}{\end{equation}}
\newcommand{\bea}{\begin{eqnarray}}
\newcommand{\eea}{\end{eqnarray}}

\newcommand{\kb}{k_{\mathrm{B}}} 
\newcommand{\mub}{\mu_{\mathrm{B}}} 

\newcommand{\dnp}{\Delta\nu_{\mathrm{p}}}
\newcommand{\dnr}{\Delta\nu_{\mathrm{r}}}

\newcommand{\sab}{\sigma_{\mathrm{Rb,b}}}
\newcommand{\sabar}{\sigma_{\mathrm{Rb,Ar}}}
\newcommand{\sabarv}{\sigma_{\mathrm{Rb,Ar}} \, v_{\mathrm{Ar}}}

\newcommand{\GGRb}{\gamma_{\mathrm{Rb}}}
\newcommand{\Isat}{I_{\mathrm{sat}}}

\newcommand{\GRb}{\Gamma_{\mathrm{Rb}}}
\newcommand{\Gjunk}{\Gamma_{\mathrm{residual}}}
\newcommand{\Ga}{\Gamma} 
\newcommand{\Gb}{\Gamma_{\mathrm{b}}} 

\newcommand{\na}{n_{\mathrm{Rb}}}

\newcommand{\Na}{N}

\newcommand{\rbs}{^{87}\mathrm{Rb}} 

\newcommand{\arf}{^{40}\mathrm{Ar}} 

\newcommand{\vrel}{\vec{v}_{\mathrm{r}}}
\newcommand{\va}{\vec{v}_{\mathrm{a}}}
\newcommand{\vb}{\vec{v}_{\mathrm{b}}}
\newcommand{\vbn}{v_{\mathrm{b}}}

\newcommand{\vrelp}{\vec{v}_{\mathrm{r}} \,'}

\newcommand{\ma}{M_{\mathrm{a}}}
\newcommand{\mb}{M_{\mathrm{b}}}

\newcommand{\dvrel}{\Delta \vec{v}_{\mathrm{r}}}
\newcommand{\dva}{\Delta \vec{v}_{\mathrm{a}}}
\newcommand{\dvb}{\Delta \vec{v}_{\mathrm{b}}}

\newcommand{\vprob}{v_{\mathrm{prob}}}

\newcommand{\tmin}{\theta_{\mathrm{min}}}
\newcommand{\uo}{U_0}

\newcommand{\nb}{n_{\mathrm{b}}}

\newcommand{\gloss}{\gamma_{\mathrm{loss}}}
\newcommand{\gcoll}{\gamma_{\mathrm{C}}}

\newcommand{\scoll}{\sigma}
\newcommand{\sloss}{\sigma_{\mathrm{loss}}}

\newcommand{\aB}{a_{\mathrm{B}}}
\newcommand{\eh}{E_\mathrm{h}}

\newcommand{\uax}{U_0^{\mathrm{axial}}}
\newcommand{\urad}{U_0^{\mathrm{radial}}}

\newcommand{\ed}{\epsilon_{\mathrm{d}}}

\begin{document}
\preprint{APS/??}
\title{Observation of quantum diffractive collisions using shallow atomic traps}
\author{David E. Fagnan$^{1}$, Jicheng Wang$^{1}$,
Chenchong Zhu$^{1}$,  Pavle Djuricanin$^{1}$, Bruce G. Klappauf $^{1}$, James L. Booth$^{2}$, and Kirk W. Madison$^{1,*}$}
\affiliation{$^1$Department of Physics \& Astronomy, University of British Columbia,\\
6224 Agricultural Road, Vancouver, BC, V6T 1Z1, Canada}
\affiliation{$^2$ Physics Department, British Columbia Institute of Technology, 3700 Willingdon Avenue, Burnaby, B.C. V5G 3H2}
\affiliation{$^*$Corresponding author: madison@phas.ubc.ca}

\date{\today}

\begin{abstract}
We present measurements and calculations of the trap loss rate for laser
cooled $\rbs$ atoms confined in either a magneto-optic or a magnetic
quadrupole trap when exposed to a room temperature background gas of Ar. We
study the loss rate as a function of trap depth and find that copious
glancing elastic collisions, which occur in the so-called quantum-diffractive regime
and impart very little energy to the trapped atoms, result in significant
differences in the loss rate for the MOT compared to a pure magnetic trap
due solely to the difference in potential depth.  This finding highlights
the importance of knowing the trap depth when attempting to infer the total
collision cross section from measurements of trap loss rates.
Moreover, this variation of trap loss rate with trap depth can be used to extract
information about the differential cross section.
\end{abstract}

\pacs{34.50.-s, 34.50.Cx, 34.80.Bm, 34.00.00, 30.00.00, 67.85.-d, 37.10.Gh}

\maketitle

\section{Introduction}

It is well known that the lifetime of a trapped ensemble of ultracold gas is limited by collisions with atoms and molecules present in the residual background vapor of an ultra-high vacuum system.  The particle loss rates and corresponding collision cross sections of laser cooled alkali metals with various species present in a background vapor have been studied since the first demonstrations of magneto-optic traps (MOTs) \cite{PhysRevLett.65.1571, PhysRevA.38.1599, Steane:92}. Small angle elastic collisions, which impart a very small amount of kinetic energy, can result in ensemble heating without immediate particle loss.  This effect has been observed experimentally \cite{PhysRevLett.70.414} and analyzed theoretically both in the classical and the purely diffractive limits \cite{cornell-1999, PhysRevA.60.R29, PhysRevA.61.033606,PhysRevA.62.063614}.  Elastic as well as inelastic collisions with a background vapor present an array of decoherence channels which can act on both the internal and external degrees of freedom of a trapped particle. Understanding the exact character of these collisions is of fundamental importance for the production of ensembles of cold atoms or molecules and their use, for example, in quantum information storage and processing applications.

While generally considered a nuisance, particle loss from a MOT induced by collisions with a background gas was recently proposed as a reliable observable to determine the total collision cross section.  Matherson \emph{et.~al} reported measurements of the absolute total collision cross section between various room temperature gases (both atomic and molecular) and laser-cooled metastable neon atoms by detecting their loss rate from a MOT \cite{Matherson07,matherson08}.  It was argued that trap loss measurements of this sort can be done with a very high precision and that they are superior to measurements of the cross sections using crossed atomic beam experiments where the determination of the absolute collision cross section is complicated by uncertainties in the absolute number of target atoms and the overlap of the beams \cite{matherson08}.  This MOT trap loss technique is similar in spirit to previous work where the measurement of absolute photoionization cross sections of atoms \cite{dinneen:1706, claessens:012706} and of absolute electron-impact ionization \cite{0295-5075-29-6-002,PhysRevLett.76.4328} were made using laser cooled atoms as a target.  In the work of Matherson \emph{et.~al}, it is claimed that their MOT loss measurements were not significantly affected by the retention of collisional products and that the cross section for trap loss was therefore equivalent to the total collision cross section \cite{matherson08}.  By contrast, Steane \emph{et.~al} argued that the collision limited trap lifetime of a Cs MOT was significantly influenced by the recapture of collisional products \cite{Steane:92}.  Similarly, Gensemer \emph{et.~al} reported dramatic variations in background collision loss rates from a Rb MOT as the trap parameters and depth were varied \cite{Gensemer97}.  In the work of Schappe \emph{et.~al} a Rb MOT was intentionally operated such that the ions produced by electron collisions would escape from the trap while atoms that were simply excited or elastically scattered were retained \cite{PhysRevLett.76.4328}.

In this work we investigate the use of trapped, laser-cooled rubidium atoms to measure the cross section for $\rbs$--$\arf$ collisions.  Motivated by the known role of small angle scattering in ensemble heating due to collisions that do not directly result in particle loss, we studied the trap loss rate produced by room-temperature background collisions as a function of trap depth at 1~K and below 10~mK using a magneto-optic and a quadrupole magnetic trap.  We find that for room temperature Ar, the retention of Rb atoms due to a finite trap depth (on the order of or larger than the energy transfer associated with quantum diffractive collisions) significantly suppresses the trap loss rate in comparison with the total collision rate.  Using model interaction potentials for the Rb--Ar complex, we confirm that the experimentally measured loss rates at various trap depths are consistent with the known long-range attractive part of the potential characterized by the $C_6$ coefficient.  These experimental and theoretical results are summarized in Fig.~\ref{fig:AvgCSVvsTrapDepthwithXERR}, and it constitutes the central result of this paper.  The strong variation of the trap loss rate as a function of the trap depth even at depths far below 1~K illustrates the very low energy-scale and abundance of quantum-diffractive collisions, which account for approximately half (exactly half for a hard core potential) of the total cross section \cite{PhysRevA.60.R29}. The conclusion is that by using a very shallow magnetic or optical dipole trap, whose depth can be easily characterized and made to approach zero, the total loss rate can be made to approach the total collision rate.  In the context of molecular beam measurements of the total collision cross section, a zero trap-depth measurement is equivalent to realizing an ideal apparatus of unlimited angular resolution \cite{Kusunoki1967}.

\section{Theory}

When a room-temperature background vapor particle encounters a trapped atom, the collision energy is typically very high.  Nevertheless, for glancing (very high impact parameter) elastic collisions, the momentum transferred and resulting scattering angle can be extremely small.  Consequently, for a trap with finite depth, collisions below a certain scattering angle do not result in trap loss.  The classical small-angle approximation predicts the cross section for loss inducing collisions increases without bound as the trap depth is reduced to zero.  This is because the inter-atomic potential is infinite in range and collisions at all ranges of impact parameter impart some, albeit small, momentum to the target atom thus producing trap loss.  However, the total cross section for loss inducing collisions can never exceed
the total collision cross section which is finite and, through the optical theorem, related to the imaginary part of the quantum mechanical scattering amplitude \cite{Landau58}.  In the limit of vanishingly small scattering angles, the associated de Broglie wavelength of the momentum transfer in the collision exceeds the classically determined impact parameter and a classical small-angle approximation is no longer valid \cite{Helbing64,anderson:2680}.  In this limit, the collisions are said to be quantum-diffractive and a full quantum-mechanical treatment of the scattering cross section is needed \cite{fagnan}.

Consider a trapped atom of velocity $\va$ and mass $\ma$ and a background particle of velocity $\vb$ and mass $\mb$. 
The initial relative velocity is $\vrel = \va - \vb$, and momentum conservation requires that the change in these velocities is related by $- \ma \dva = \mb \dvb$ and
$\dva = \frac{\mu}{\ma} \dvrel$, where the reduced mass is $\mu = \frac{\ma \mb}{\ma + \mb}$.
In addition, energy conservation requires that for elastic collisions
$|\dvrel|^2 = 2 |\vrel|^2 (1-\cos \theta)$, where $\theta$ is the collision angle between the initial ($\vrel$) and final ($\vrelp$) relative velocities.
After the collision, the trapped atom kinetic energy will have changed by an amount
$\Delta E = \frac{\ma}{2} \left[(\va + \dva)^2 - \va^2\right]  =  \frac{\ma}{2} \left[ 2 \va \cdot \dva + |\dva|^2 \right]$.  If this change in kinetic energy exceeds the trap depth $\uo$, loss will occur.  In the limit that the trapped particle has a negligible initial kinetic energy ($\va \ll \dva$), we have
that $\Delta E \simeq  \frac{\ma}{2} |\dva|^2$ and
\be
\Delta E \simeq  \frac{\mu^2}{\ma} |\vrel|^2 (1-\cos \theta).
\ee
Therefore, if the collision angle exceeds the minimum angle
\be
\tmin = \arccos \left(1- \frac{\ma \uo}{\mu^2 |\vrel|^2} \right),
\ee
then trap loss will occur.  The rate at which background particles are scattered from a single trapped atom into a solid angle $d\Omega$ is $\nb |\vrel| (d \sigma / d \Omega) d \Omega$, where  $\nb$ is the density of background particles and $(d \sigma / d \Omega)$ is the differential scattering cross section.  Given this, we can estimate the loss rate from a trap of depth $\uo$ as
\be
\gloss = \nb \vprob \sloss = \nb \vprob \int_{\tmin(\vprob)}^{\pi} (d \sigma / d \Omega) d \Omega.
\ee
where the relative collision velocity is assumed to be determined by the most probable velocity for the background particles, $|\vrel| =\vprob$.  This loss rate is always smaller than the estimated total collision rate, $\gcoll = \nb \vprob \scoll $, since $\tmin > 0$.  The differential cross section is related to the quantum mechanical scattering amplitude,
$d \sigma/d\Omega = |f(k, \theta)|^2$, and depends explicitly on the collision wave vector, $k = \mu |\vrel| / \hbar$.  It is assumed that the interaction potential is central and therefore there is no azimuthal angular dependence.  For a beam of incident scattering particles with wave vector $k$, the cross section for loss inducing collisions from a trap of depth $\uo$ is
\be
\sloss(k) = \int_{\tmin(\hbar k / \mu)}^{\pi} 2 \pi \sin \theta |f(k,\theta)|^2 d\theta.
\label{eq:sigmaloss}
\ee
This expression is equivalent to the total cross section measured by a molecular beam apparatus with finite angular resolution limited to $\tmin$ \cite{Kusunoki1967}.

To compute the loss rate induced by collisions with a background gas at temperature $T$, we need to average over the Maxwell-Boltzman distribution of incident collision wave vectors.  Assuming the trapped atom velocity is negligible, we have that $\vrel \simeq \vb$ and we can compute the velocity averaged loss rate
using $k= \mu |\vb| / \hbar$ and find $\langle \gloss \rangle = \nb \langle v \sigma \rangle$ where
\be
\langle v\sigma \rangle = \left(\frac{\mb}{2 \pi \kb T}\right)^{3/2} \int_{0}^{\infty} 4 \pi \, \sloss(k) \, \vbn^3 e^{- \mb \vbn^2/2 \kb T} d\vbn.
\label{eq:lossrateavg}
\ee

In order to evaluate this expression, we need to determine the scattering amplitude, $f(k,\theta)$, given the inter-atomic potential between atom a and b.  The asymptotic form of the 2-body scattering wave function is the sum of an incident plane wave and a scattered spherical wave
$\psi(\mathbf{r})|_{r \rightarrow \infty}  = A \left( e^{ikz} + f(k,\theta) \frac{e^{ikr}}{r}\right).$  Given a central potential, $\psi$ and $f$ can be expanded in terms of the Legendre polynomials,
\bea
\psi_k(r, \theta) & = & \sum_{l=0}^{\infty} R_l(k,r) P_l(\cos \theta),\\
f(k, \theta) & = & \sum_{l=0}^{\infty} f_l(k) P_l(\cos \theta).
\label{eq:LegExp}
\eea
This is an expansion into \emph{partial waves} of angular momenta $l$.  For sufficiently large $r$, the potential is negligible, and the radial functions asymptotically approach the form for a free particle
\be
R_l(k,r)|_{r \rightarrow \infty} = B_l j_l(kr) + C_l n_l(kr),
\label{eq:radialassymptotic}
\ee
where $j_l$ and $n_l$ are the spherical Bessel and Neumann functions.  The
coefficients $B_l$ and $C_l$ can be written as
$B_l = A_l \cos \delta_l$ and $C_l = - A_l \sin \delta_l$,
where $A_l = (2 l +1 ) i^l e^{i \delta_l}$
and where $\delta_l = \arctan(-C_l / B_l)$ is the \emph{phase shift} of the $l^{\mathrm{th}}$ partial wave \cite{Child,Markovic}.  These phase shifts also determine the scattering amplitude 
\be
f(k,\theta) = \frac{1}{k} \sum_{l=0}^{\infty}(2l+1) e^{i \delta_l} \sin \delta_l P_l(\cos \theta), \label{eq:scattampl}\\
\ee

The determination of the scattering amplitude and resultant collision cross section therefore requires finding the partial wave phase shifts, and they are found by numerical integration of the radial 
Schr\"odinger equation.  We write the radial equation for the $l^{\mathrm{th}}$ partial wave as
\be
\left[ \frac{d^2}{dr^2} + W_l(r) \right]\psi_l(r) = 0
\ee
where 
\bea
W_l(r) & = & k^2 - \frac{2\mu}{\hbar^2} V(r) - \frac{l(l+1)}{r^2},\\
\psi_l(r) & = & kr R_l(r),
\eea
and $V(r)$ is the inter-atomic potential.  The solution to the radial equation for each partial wave $l$ is then independently computed using the log-derivative method \cite{Johnson:445}.  The integration of the radial equation starts from a position deep inside the classically forbidden region and ends at a point in the asymptotic regime, beyond which the contribution becomes insignificant.  The phase shift is found by matching this numerical solution in the asymptotic region to the asymptotic form, Eq.~\ref{eq:radialassymptotic}, and by using $\frac{-C_l}{B_l} = \tan \delta_l$.

\begin{figure}[h]
\centering
\includegraphics[width=3in]{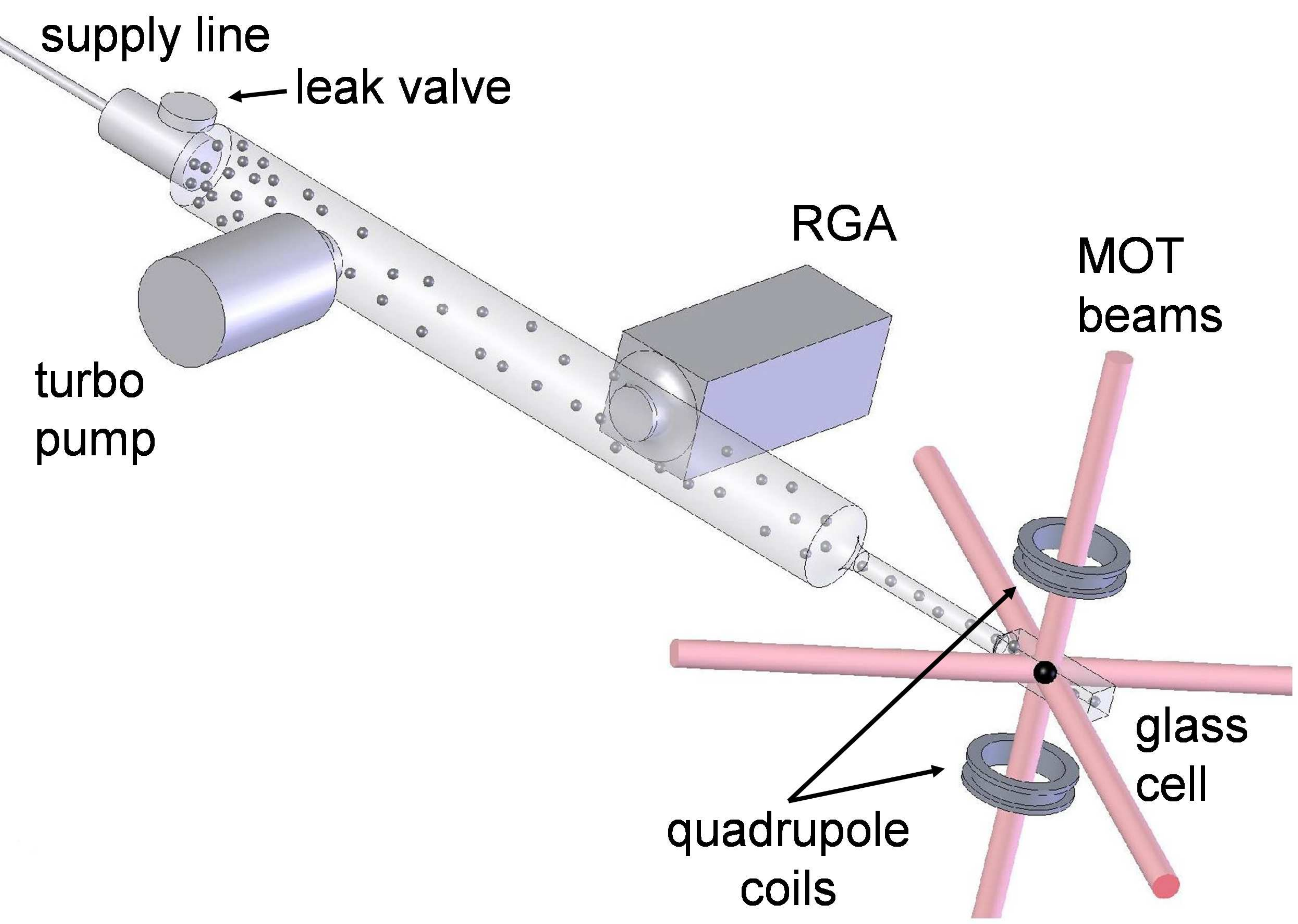}
\caption{(Color online) Schematic of the vacuum system. The variable leak valve allows Ar gas to be introduced and the residual gas analyzer (RGA) is used to measure the pressure and composition of the background vapor. The laser cooled and trapped $\rbs$ atoms undergo collisions with the background gas which may or may not result in trap loss.}
\label{fig:ExpSetup}
\end{figure}

The resultant values for partial wave phase shifts $\delta_l$ depend on the incident
wave vector magnitude $k = \mu |\vrel| / \hbar$.  Given the results for a particular collision wave vector, the scattering amplitude can be constructed using Eq.~\ref{eq:scattampl} and inserted into Eq.~\ref{eq:sigmaloss} to compute the total cross section for loss inducing collisions at that collision velocity.  Repeating this procedure for a set of wave vectors chosen from a Maxwell-Boltzman distribution, the velocity averaged rate of loss inducing collisions from a background gas at temperature $T$ (Eq.~\ref{eq:lossrateavg}) is evaluated.  In each numerical simulation, the convergence of these results is checked both as a function of the integration range and step size and as a function of the number of partial waves included.  It is found empirically that the results (for room temperature collisions) have converged to an accuracy of better than 1\% when partial waves up to $l=500$ are included (higher velocities require more partial waves) and given the integration is performed with a step size of 0.005~$\aB$ ($\aB = 0.529177$~\AA) from 1~$\aB$, deep inside the classically forbidden region, to 150~$\aB$, well in the the asymptotic regime.

\section{Experimental Setup}

Our apparatus consists of a magneto-optic trap (MOT) which collects and cools a few million rubidium atoms from a room-temperature Rb vapor of below $10^{-8}$~torr.  The atomic vapor is maintained by periodically heating an Alvatec rubidium source.  The laser system used for $^{87}$Rb laser cooling is composed of grating-stabilized and injection-seeded diode lasers and has been described previously \cite{Ladouceur:09}.  After amplification, we have a total of 20~mW of pump or cooling light (driving transitions from $|F=2\rangle \rightarrow |F'=3\rangle$) and 4~mW of repump light (driving transitions from $|F=1\rangle \rightarrow |F'=2\rangle$) both operating on the $5^2S_{1/2} \rightarrow 5^2P_{3/2}$ transition. This light is combined and then expanded to a $1/e^2$ diameter of 7~mm, prepared with the correct polarization and introduced into the MOT cell (a borosilicate glass box) along the three mutually orthogonal axes in a retro-reflection configuration.  We typically operate with an axial gradient of $b'=23.1(0.7)$~G\,cm$^{-1}$ and with pump and repump detunings of $\dnp= -2 \GGRb$ and $\dnr=0$, where $\GGRb=2 \pi \times 6.07$~MHz is the natural linewidth of the $5^2S_{1/2} \rightarrow 5^2P_{3/2}$ transition.  The total pump intensity is $I = 5.6 \, \Isat$ (where $\Isat=3.58$mW\,cm$^{-2}$ for an isotropic pump field with equal components in all three possible polarizations \cite{Steck08}).  Based on previous measurements of the escape probability as a function of energy for a Rb MOT \cite{PhysRevA.54.R1030} and studies of the dependence of the capture velocity on intensity \cite{Gensemer97,PhysRevA.62.013404}, we estimate the MOT trap depth in this work to be $800 \pm 300$~mK.  The number of trapped atoms in the MOT is determined by measuring the emitted resonance fluorescence.

\begin{figure}[h]
\centering
\includegraphics[width=3in]{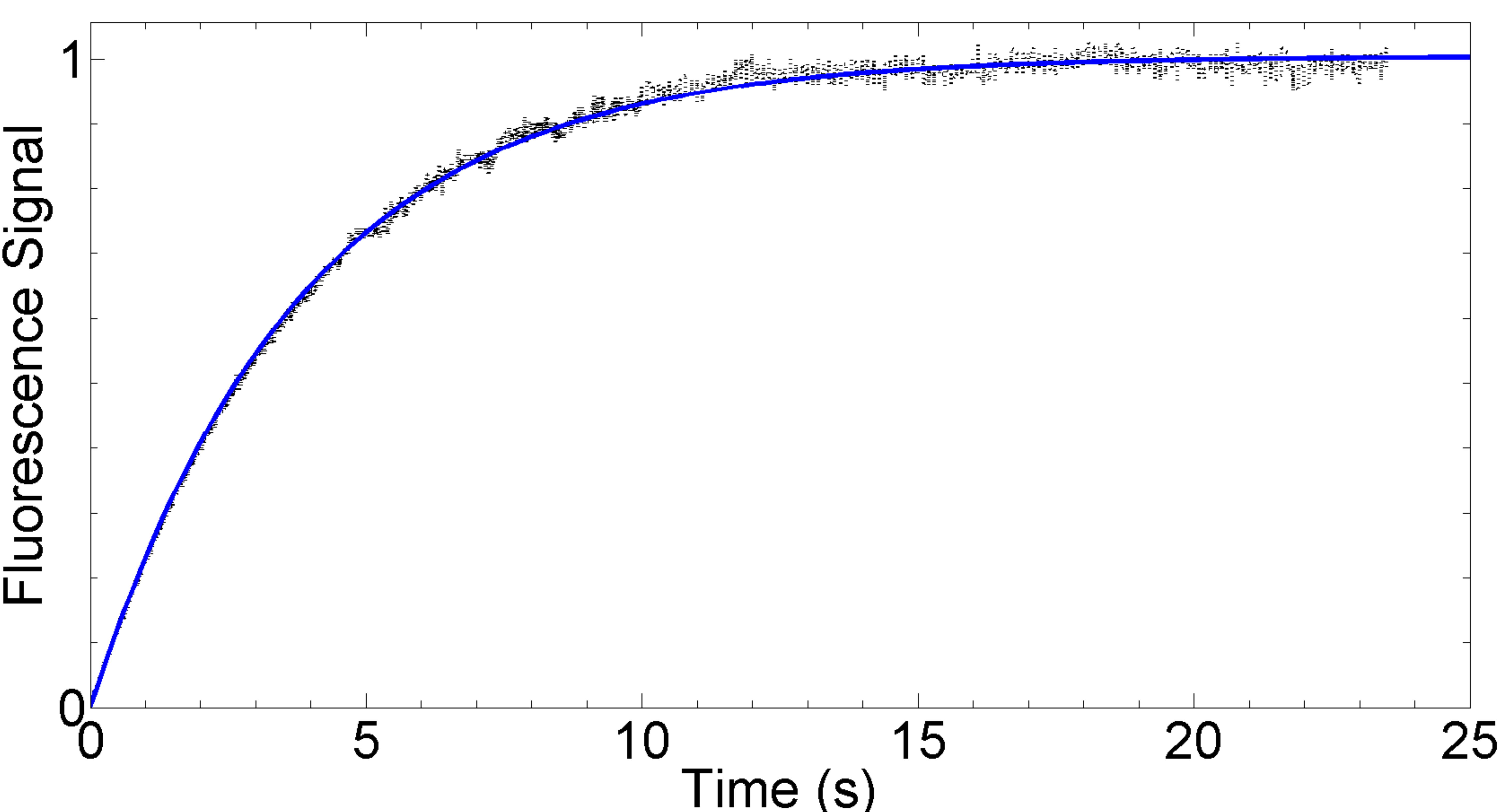}
\caption{(Color online) MOT Fluorescence as a function of time fit to the rate equation given in Eq.~\ref{eq:RateEquation} with $\beta=0$.}
\label{fig:MOTLoadingCurve}
\end{figure}

$\rbs$ atoms are cooled, optically pumped into either the $|F=1\rangle$ or $|F=2\rangle$ hyperfine ground state and then transferred from the MOT into a quadrupole magnetic trap.  For this purpose, the MOT anti-helmholtz coil pair is used and provides an axial gradient of up to $b'=670(20)$~G\,cm$^{-1}$.  The intrinsic loss rate in this quadrupole magnetic trap due to Majorana spin flips is negligible at the temperatures and densities in this experiment \cite{PhysRevLett.74.3352}.  The atom number remaining in the magnetic trap after some hold time is measured by rapidly returning the magnetic field to an axial gradient of $b'=23.1(0.7)$~G\,cm$^{-1}$, used for the MOT, switching the MOT lasers on and detecting the fluorescence of the atoms which are recaptured in the MOT.  Within a millisecond, the atoms are laser cooled and recollected in the MOT and the resonance fluorescence is monitored during the first 100 ms after recapture. The measured atom number is determined from this fluorescence and is corrected for the number of atoms captured from the residual $\rbs$ background vapor during the measurement time.  Since the trap loss rates reported here are determined from the ratios of atom numbers remaining in the trap after various hold times, the only requirement for a precise determination of the loss rate is that the measured fluorescence be linear in the number of trapped atoms.  Since the optical density is well below 0.5 for all of the measurements reported here, this is the case.

\begin{figure}[h]
\centering
\includegraphics[width=3in]{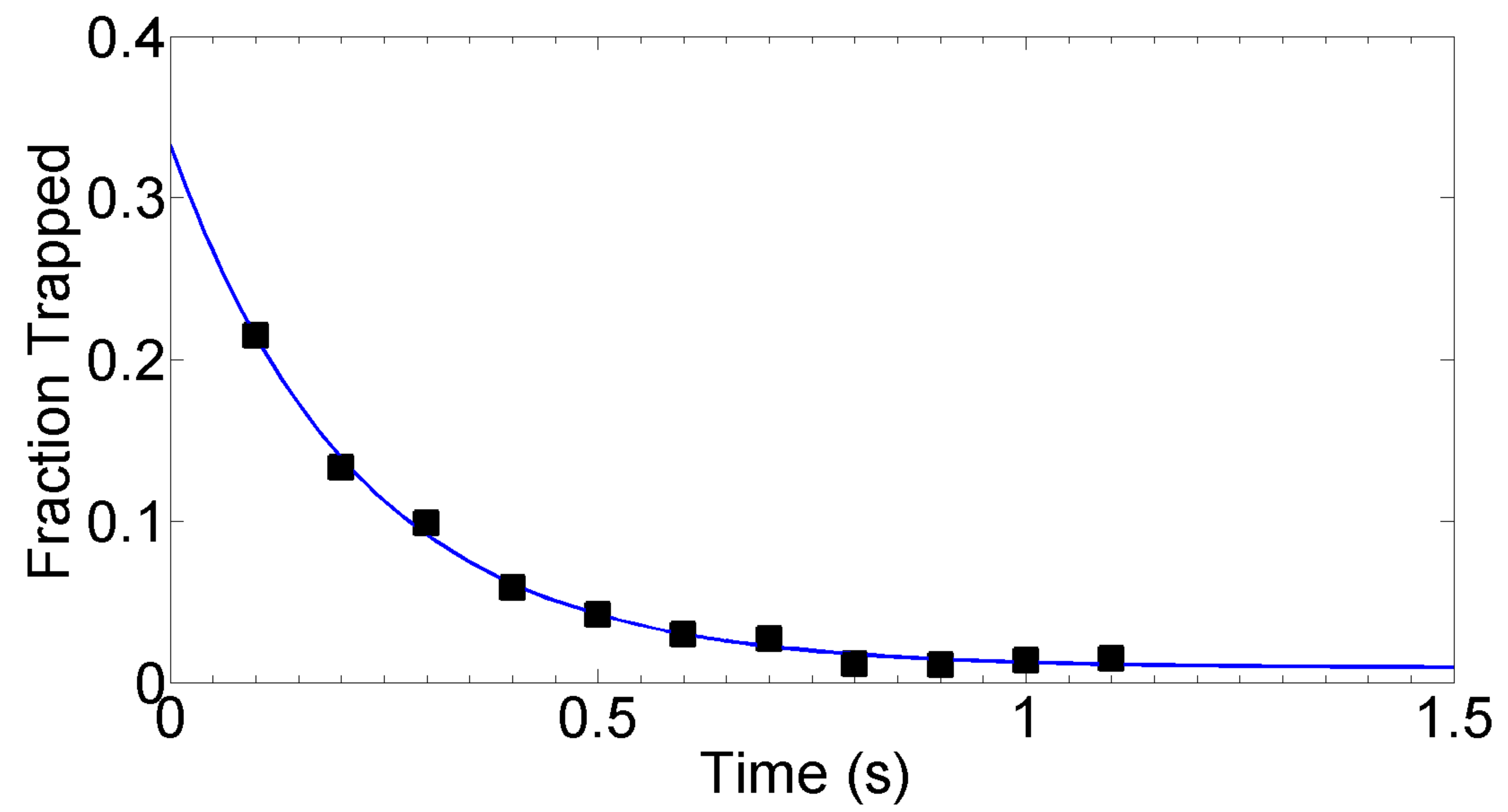}
\caption{(Color online) Fraction of atoms retained in a quadrupole magnetic trap as a function of time fit to the rate equation given in Eq.~\ref{eq:RateEquation} with $R=0$ and $\beta=0.$
}
\label{fig:MagTrapDecay}
\end{figure}

Additional room-temperature Ar gas is introduced into the vacuum envelope through a variable leak valve.  This gas is continuously pumped away by a turbo-molecular pump, and the resulting equilibrium partial pressure is controlled by the leak rate of the value and measured using a residual gas analyzer (SRS RGA 200) as shown in Fig.~\ref{fig:ExpSetup}.  The additional trap loss rate induced by this new gas as a function of its pressure is measured.

Although an accurate determination of the exact cross section using this trap loss method would require careful calibration of the residual gas pressure, the focus of this work is to illustrate the dependence of the collision cross section on trap depth. Therefore this work requires only that the the ratio of cross sections at different trap depths be precise. In order to ensure this, and to mitigate the effect of the uncertainty in the absolute pressure, the data were collected as follows: the Ar pressure was first held fixed and the loss rates for the MOT and magnetic trap at various depths were measured. The pressure was then adjusted and the procedure was repeated. This ensured that although the absolute pressure reading may be in error, the loss rates in traps of varying trap depth were all measured at the same pressures.  The ratio of cross sections between traps of various depth is therefore precise to within a statistical error of less than 10\%.

In the MOT, the atom number loading dynamics in the presence of a background vapour
is modelled by the following rate equation \cite{Weiner99},
\be
\label{eq:RateEquation}
\frac{d\Na}{dt} =  \mathrm{R} - \Ga\Na -  \beta \int \na^2 \, dV
\ee
where $\Na$ is the trapped atom number, $\mathrm{R}$ is the loading rate, $\Ga$ is the loss rate due to collisions between the trapped rubidium atoms and the room-temperature background gas, and $\beta$ is the combined density dependent 2-body loss coefficient due to cold collisions within the MOT.  Here 3-body and higher order collisions are neglected.  The density profile of the cold, trapped atoms is given by $\na$, and these density dependent losses include light-assisted collisional losses (such as radiative escape and fine structure changing collisions) and hyperfine structure changing collisions.  For sufficiently small atom numbers, the last term is negligible compared to the second \cite{Ladouceur:09}.  In the case of the pure magnetic quadrupole trap, the trapped number is well described by this model with $R=0$.  In the magnetic trap, the confinement is much weaker than the MOT and the trap volume (up to 5~mm in radius compared to the MOT cloud radius of typically 150~$\mu$m) is larger by a factor of more than $10^{4}$, making the density dependent collisional losses even smaller.  In addition, light assisted collisions (which dominate the $\beta$ term for the MOT) are no longer present, and only hyperfine changing collisions and spin relaxation collisions (from magnetic dipole-dipole coupling) remain.  In any case, the data presented here are taken in the regime where these 2-body loss processes are completely dominated by the $\Ga N$ term.

\section{Results}

Figures~\ref{fig:MOTLoadingCurve} and \ref{fig:MagTrapDecay} show the loading dynamics of the MOT and the atom loss from a magnetic trap fit to Eq.~\ref{eq:RateEquation}.  From these fits, $\Ga$ is extracted. The loss rate due to collisions with the room temperature gas includes a contribution from three different terms
\be
\label{eq:LossRate}
\Ga = \GRb+\Gjunk+\Gb.
\ee
$\GRb$ is the loss rate due to collisions with room-temperature Rb atoms, and $\Gjunk$ is the loss rate due to collisions with unwanted residual gas in the ultra-high vacuum.  $\Gb$ is the loss rate due to collisions with gas particles intentionally introduced into the vacuum system and is given by $\Gb = \nb  \langle \sab \vbn  \rangle$ where $\langle \sab \vbn  \rangle$ is the velocity averaged collision cross section of interest (Eq.~\ref{eq:lossrateavg}).

As the controlled background gas density $\nb$ is increased, the total loss rate, $\Ga$, increases with slope $\langle \sab \vbn \rangle$.  In general, all three of the loss rates contributing to $\Ga$ ($\GRb$, $\Gb$ and $\Gjunk$) can be different in the MOT than in a magnetic trap.  This is due to the difference in trap depths and to the fact that some of the atoms in the MOT are in an electronic excited state during a collision (on the order of 12\% for our MOT parameters) which will, in general, have a different interaction potential and corresponding cross section with the colliding particles than do ground state Rb atoms. The case of excited state Rb atoms colliding with identical ground state Rb atoms is an extreme example.  The leading order in Rb--Rb ground state interactions is $C_6/r^6$ whereas it is $C_3/r^3$ for Rb--Rb* collisions. In this way, a measurement of magnetic trap loss is less complicated to interpret since it only involves elastic collisions of ground state atoms.

\begin{figure}[h]
\centering
\includegraphics[width=3in]{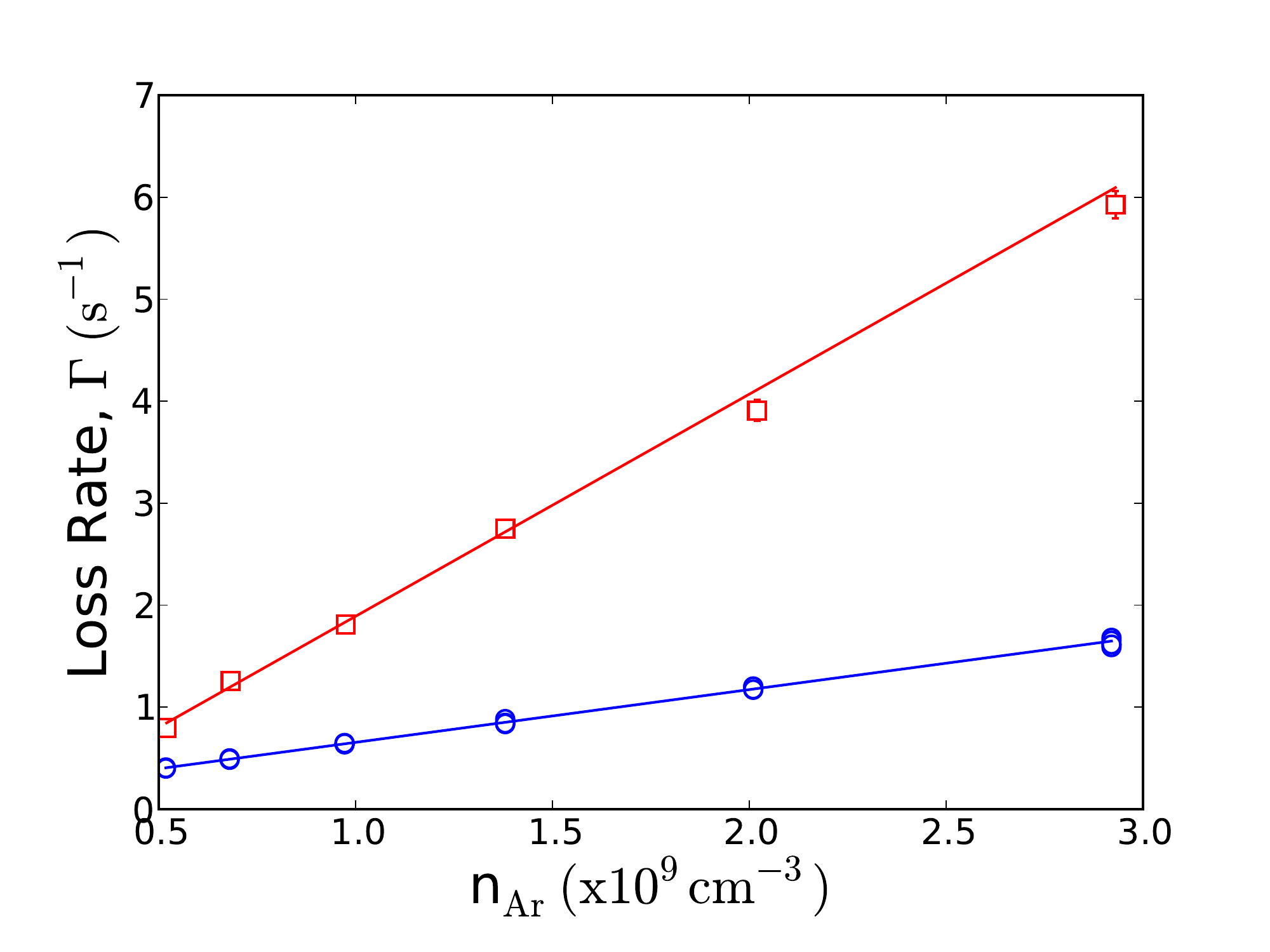}
\caption{(Color online) Loss rate of trapped $\rbs$ versus room temperature argon gas density.  The results are fit to a line and the slope provides the value of the velocity averaged collision rate $\langle \sabarv   \rangle$.  The MOT (circles) exhibits the smallest cross section due to a much larger effective trap depth ($\sim$1~K) than the magnetic trap (squares) where the $|F=1,m_F=-1\rangle$ state was confined with $b'=79$~G\,cm$^{-1}$ resulting in a trap depth of 0.64~mK. The vertical error bars, most of which are smaller than the symbols, represent the statistical uncertainty in the loss rate based on fits similar to those shown in Figs.~\ref{fig:MOTLoadingCurve} and \ref{fig:MagTrapDecay}.
}
\label{fig:LossRate}
\end{figure}

Figure~\ref{fig:LossRate} shows the measured loss rate of trapped $\rbs$ versus argon density in both a MOT and a quadrupole magnetic trap. These data are well fit by a line and the slopes, which provide the value of the velocity averaged collision rate $\langle \sabarv  \rangle$, are listed in table~\ref{tab:lossrateslopes}.  The MOT exhibits the smallest cross section (i.e. slope) due to a much larger trap depth ($\sim$1~K) than the magnetic trap ($\sim$1~mK).  As the argon density and associated background collision loss rate are increased, the steady state MOT number is smaller as are any density dependent losses (our MOT is operating in the small-particle range where the density increases linearly with atom number \cite{Ladouceur:09}).  If the density dependent losses were significant, one would therefore expect the loss rate to vary non-linearly with argon gas density.  The MOT loss rate is observed to vary linearly with the argon density confirming that the combined density dependent 2-body losses due to cold collisions within the MOT are negligible under our experimental conditions.

Before loading the pure magnetic trap, hyperfine optical pumping is employed to populate and confine either the $|F=2\rangle$ or the $|F=1\rangle$ state.  For small magnetic fields where the electronic and nuclear magnetic moments are strongly coupled by the hyperfine interactions, $F$ is a good quantum number and the $|F=1, m_F=-1\rangle$ and $|F=2, m_F=1,2\rangle$ states are the magnetically trappable, weak-field seeking states.  Also at small magnetic fields, the $|F=1, m_F=-1\rangle$ and $|F=2, m_F=1\rangle$ states have essentially identical magnetic moments (differing  by less than 0.0004\% at zero magnetic field) while the $|F=2, m_F=2\rangle$ state has twice the magnetic moment of the others and therefore experiences a larger trap depth for a given magnetic field gradient.  The quadrupole coils are oriented with the axis of symmetry along the direction of gravity (with an axial gradient of up to $b'=670(20)$~G\,cm$^{-1}$ and a radial gradient of $b'/2$).  For small fields, the magnetic trap potential energy along the axial direction ($z$) is $\uax = \alpha b' |z| + mgz$, where $\alpha = \mub g_F m_F$ is the magnetic moment of the state, $\mub \simeq h \times 1.4$~MHz/G is the Bohr magneton, and $g_F = 1/2 (-1/2)$ for the $|F=2\rangle$ ($|F=1\rangle$) state.  Along the radial direction, the trap potential energy is, for small magnetic fields, $\urad = \alpha \frac{b'}{2} r$.  

\begin{figure}[h]
\centering
\includegraphics[width=3in]{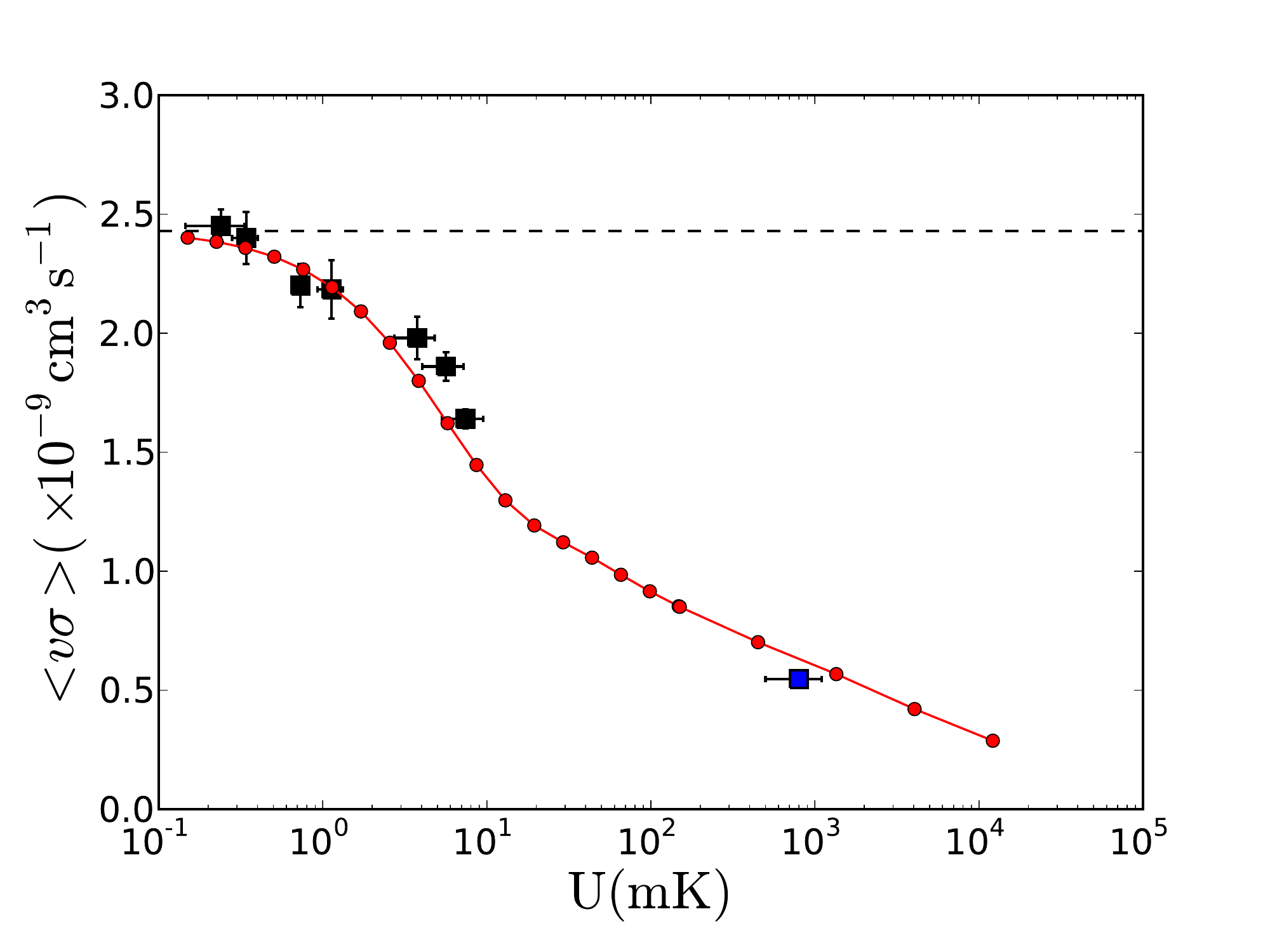}
\caption{(Color online) The experimentally measured (squares) and theoretically computed (circles) loss rate slope,  $\langle \sabarv \rangle$, as a function of trap depth for $\rbs$--$\arf$ collisions.  The line is only a guide to the eye.  The data point at 1~K was obtained using a MOT and the data below 10~mK were obtained with a quadrupole magnetic trap.  These values were extracted from the data shown in Fig~\ref{fig:LossRate}. The horizontal error bar for the MOT data point represents an estimated trap depth range based on the work of Hoffmann \emph{et.~al} \cite{PhysRevA.54.R1030}.  The error bars for the data below 10~mK represent the radial and axial magnetic trap depths. The horizontal dotted line indicates the zero trap depth limit for $\langle \sabarv \rangle$, corresponding to the total collision cross section.
}
\label{fig:AvgCSVvsTrapDepthwithXERR}
\end{figure}

For large magnetic fields, the electronic and nuclear spin magnetic moments begin to become decoupled and the Zeeman shift is non-linear.  The trap depths reported (Table~\ref{tab:lossrateslopes} and Fig.~\ref{fig:AvgCSVvsTrapDepthwithXERR}) were evaluated by diagonalizing the full Hamiltonian including the hyperfine and Zeeman interactions.   The magnetic trap depth is limited by collisions with the cell walls which are 5~mm from the magnetic trap center along the radial and axial directions.  Since particles can escape along either the axial or radial directions, a trap depth error bar is provided in Table~\ref{tab:lossrateslopes} and Fig.~\ref{fig:AvgCSVvsTrapDepthwithXERR} indicating the depths along these directions. Typically, the trap depth is limited by radial trajectories; however, at axial gradients below $b' < \frac{2mg}{\alpha}$ gravity limits the trap depth along the $z$ direction.  In addition, below a critical gradient of $b'<\frac{mg}{\alpha}=30.5$~G\,cm$^{-1}$, only the $|F=2, m_F=2\rangle$ is supported against gravity.  This feature can be used to ``clean" the $|F=2\rangle$ manifold removing all atoms in the other low-field seeking states.  As discussed above, for the lower hyperfine state, only the $|F=1,m_F=-1\rangle$ state is magnetically trapped, while for the upper hyperfine state, due to the non-linear Zeeman shift, there are three states (which correlate to the $|F=2,m_F=2,1,0\rangle$ states at zero magnetic field) which can be trapped magnetically, and each state has a different trap depth.  To simplify the interpretation of the data, we focus on the loss rate of the $|F=1,m_F=-1\rangle$ state.  In addition, we study trap loss for the upper hyperfine state at a field gradient of 23.7~G\,cm$^{-1}$.  At this gradient only the $|F=2, m_F=2\rangle$ state is trapped and this ensures a single trap depth for the trapped atomic ensemble.

\begin{table}
\caption{Measurements of the trap loss rate slope for $\rbs$ in the presence of room temperature $\arf$ gas listed in order of increasing trap depth.  These results are compared with the theoretical computed trap loss rate slope shown in Fig.~\ref{fig:AvgCSVvsTrapDepthwithXERR}.  The minimum and maximum values within the uncertainty in the magnetic trap depth indicate the depth along the radial and axial directions while the uncertainty in the MOT trap depth represents the estimated trap depth range based on the work of Hoffmann \emph{et.~al} \cite{PhysRevA.54.R1030}.
}
\begin{ruledtabular}
\begin{tabular}{c|c|c|c}
Trapped State & $b'$  & Trap Depth & $\langle \sabarv \rangle$\\
$|F, m_F\rangle$ & (G\,cm$^{-1}$) & (mK) & ($\times 10^{-9}$cm$^3$\,s$^{-1})$\\
\hline
$|1,-1\rangle$ & 39.5 & 0.24 (0.09) & 2.45 (0.07)\\
\hline
$|2,+2\rangle$ & 23.7 & 0.34 (0.06) & 2.40 (0.11)\\
\hline
$|1,-1\rangle$ & 79.0 & 0.73 (0.07) & 2.20 (0.09)\\
\hline
$|1,-1\rangle$ & 111 & 1.13 (0.20) & 2.18 (0.12)\\
\hline
$|1,-1\rangle$ & 334 & 3.77 (1.04) & 1.98 (0.09)\\
\hline
$|1,-1\rangle$ & 501 & 5.64 (1.59) & 1.86 (0.06)\\
\hline
$|1,-1\rangle$ & 668 & 7.41 (2.09) & 1.64 (0.04)\\
\hline
MOT & 23.7 & 800 (300) & 0.54 (0.02)\\
\end{tabular}
\end{ruledtabular}
\label{tab:lossrateslopes}
\end{table}

\begin{figure}[h]
\centering
\includegraphics[width=3in]{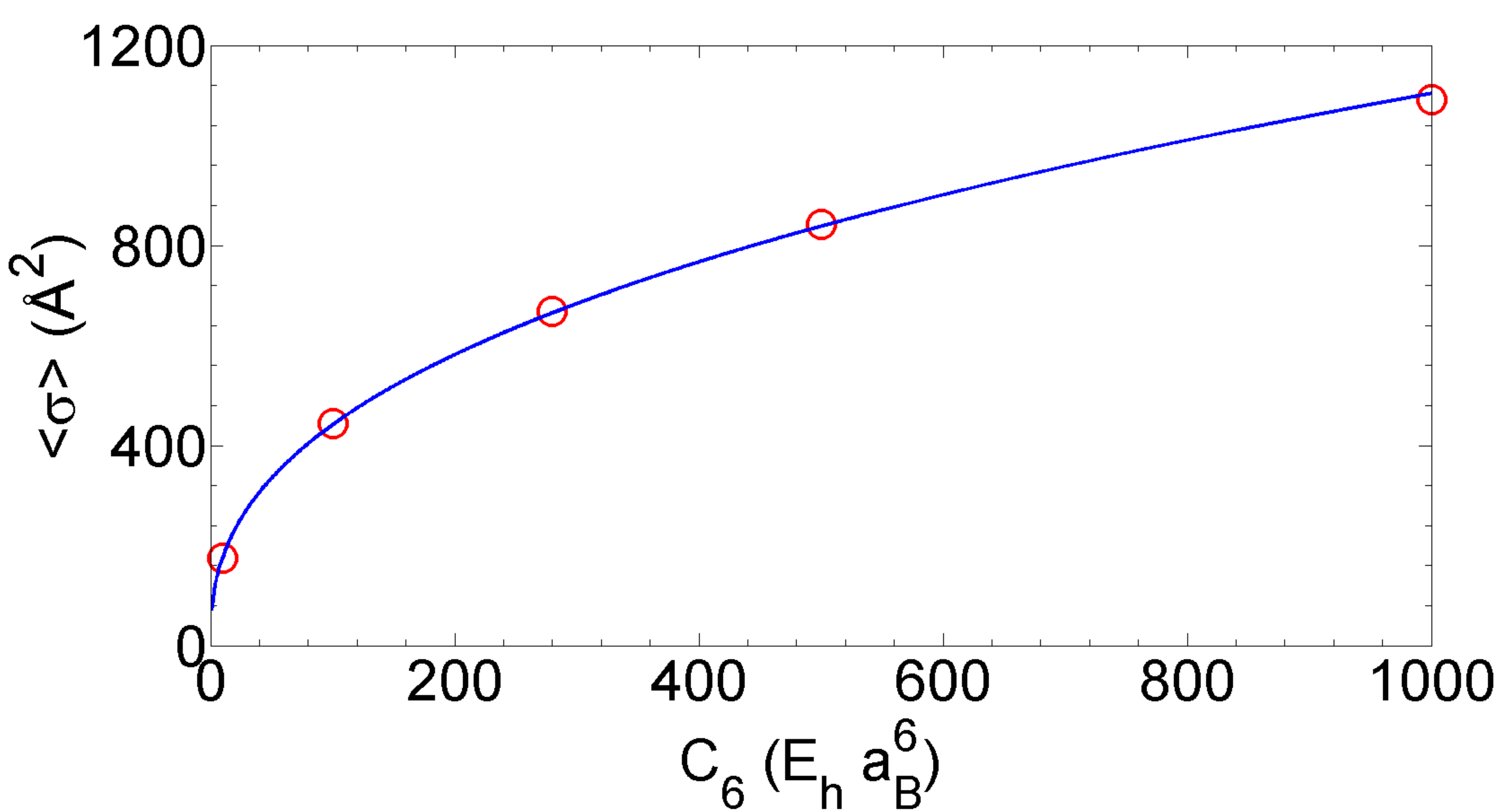}
\caption{(Color online) The theoretically computed total cross section for room temperature $\rbs$--$\arf$ collisions (averaged over a Maxwell-Boltzmann distribution at 295~K) given a Lennard-Jones inter-atomic potential as a function of the $C_6$ coefficient ($C_{12}=8.6 \times 10^7~\eh \aB^{12}$). The numerical results are fit to a power law, $\sigma = 70 \AA^2 \times (C_6)^{0.40} $, in agreement with the approximate dependence of 2/5.
}
\label{fig:AvgCSVsC6}
\end{figure}

In this analysis, the inter-atomic interaction is modeled by a Lennard-Jones potential, $V(r) = C_{12}/r^{12}-C_6/r^6$, where the value $C_6=280~\eh \aB^6$ is used for the Rb-Ar interaction ($\eh = 4.35974 \times 10^{-18}$~J).  This coefficient is known from both optical and atomic beam measurements \cite{rosin:373,dalgarno:479,mahan:950}. $C_{12}=8.6 \times 10^7~\eh \aB^{12}$ is chosen so that the resulting potential has a minimum of 50~cm$^{-1}$ below zero.  This is the order of magnitude expectation for the Ar-Rb complex \cite{romanexplain}.  However, this choice is somewhat arbitrary, since for the collision energies involved in this experiment, the exact value of $C_{12}$ has a negligible effect on the total cross section.  For a long-range potential of the form $V(r) = - C_n / r^n$, where $n>2$, it can be shown \cite{Landau58} that the cross section can be expressed as
\be
\sigma(k) = A \left( \frac{C_n}{\hbar v}\right)^{2/(n-1)},
\label{eq:sigmaVsCn}
\ee
where $A$ is a constant and $v$ is the collision velocity.  This expression is approximate and only valid for sufficiently large values of the collision wave vector $\mu C_n k^{n-2}/\hbar^2 \gg 1$ (satisfied for all Ar--Rb collision velocities above 1 m/s for $C_6=280~\eh \aB^6$) and for potentials which have the form specified for distances beyond $r_c \sim (C_n/\hbar v)^{1/(n-1)}$.  Given the known value for $C_6$ and our choice for $C_{12}$, the Lennard-Jones potential deviates from a pure $C_6/r^6$ potential by less than 3\% at distances beyond $r_c$, defined for Ar--Rb collision velocities of 1000 m/s.  The deviation is even smaller for lower collision velocities since $r_c$ is larger.  The result is that for collisions with a room temperature gas, the cross section is insensitive to the exact value of $C_{12}$.  Of course, for much higher velocities, the exact short range behavior of the inter-atomic potential begins to matter, and at velocities much lower than 1~m/s, low partial wave collisions are dominant and the above approximate expression is invalid.

As a check of the numerical calculations, the computed cross section is compared with the predicted power law dependence on $C_6$, and an exponent of 0.40 is found, in agreement with the expected 2/5 dependence from Eq.~\ref{eq:sigmaVsCn}.  The results are shown in Fig.~\ref{fig:AvgCSVsC6}.

\begin{figure}[h]
\centering
\includegraphics[width=3in]{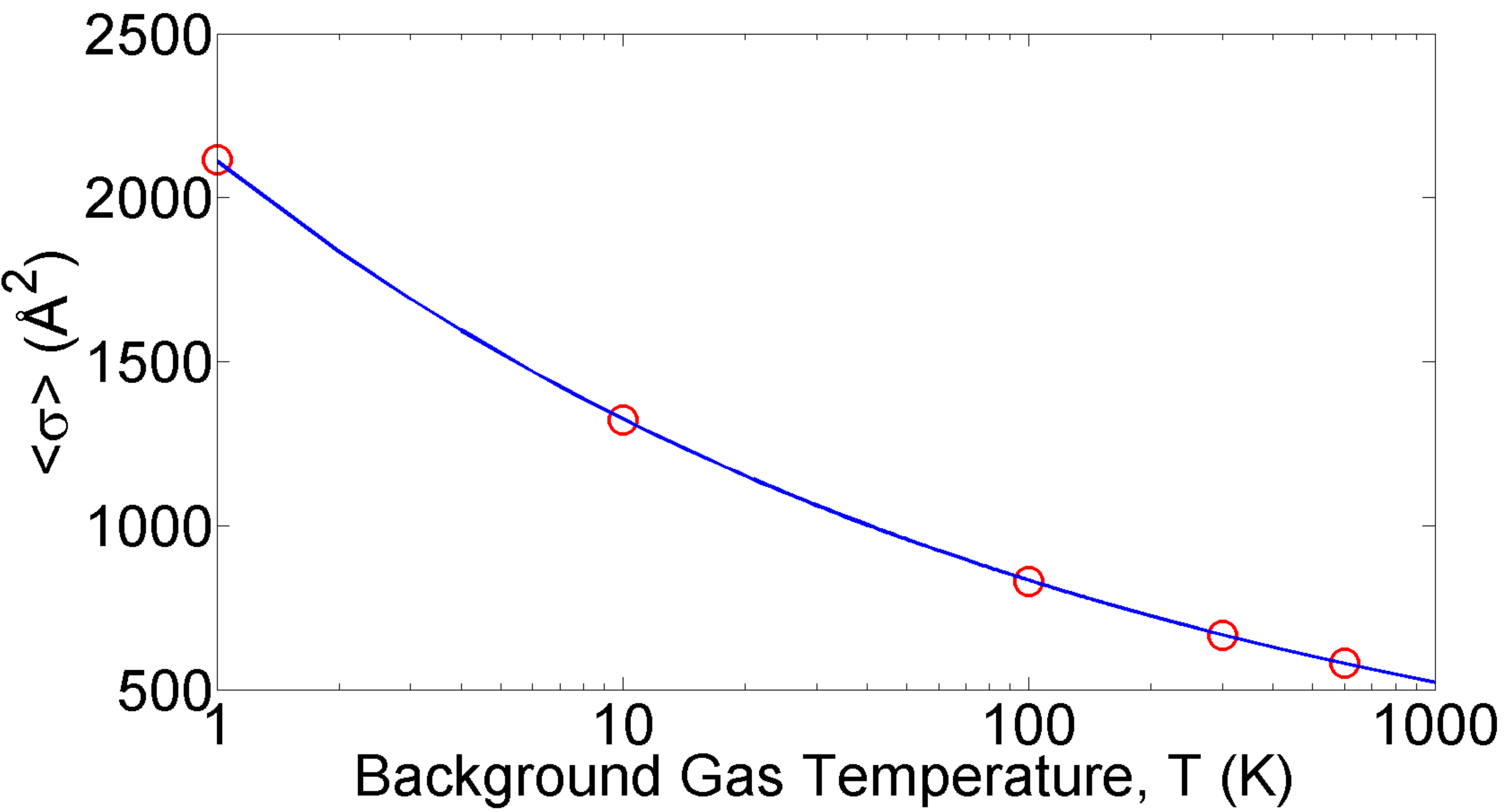}
\caption{(Color online) The theoretically computed total cross section for $\rbs$--$\arf$ collisions averaged over a Maxwell-Boltzmann distribution at various temperatures given a Lennard-Jones inter-atomic potential with $C_6=280~\eh \aB^6$ and $C_{12}=8.6 \times 10^7~\eh \aB^{12}$.  The results are fit to a power law, $\sigma = 2100 \AA^2 \times (T)^{-0.202}$, in agreement with the approximate expected power dependence of -1/5, corresponding to the power dependence of -2/5 on the collision velocity.
}
\label{fig:AvgCSVsTemp}
\end{figure}

The total cross section for $\rbs$--$\arf$ collisions averaged over a Maxwell-Boltzmann distribution,
$\langle \sabar \rangle$, is also computed at various Ar temperatures and shown in 
Fig.~\ref{fig:AvgCSVsTemp}.  These results also agree with the predictions of Eq.~\ref{eq:sigmaVsCn}.  In addition, they are consistent with previous atomic beam measurements of Ar--Rb collisions.  In that work, the attenuation of a 283~K Ar beam traveling through a 590~K Rb vapor was measured and a cross section of 572 $\AA^2$ was found \cite{rosin:373}.  Although the average beam velocity was not specified, we estimate that the relative collision speed in that measurement corresponds to a stationary Rb atom exposed to a $T > 400$~K Ar gas.  Our calculations predict a cross section of 
626~$\AA^2$ at $T=400$~K,  within 10\% of the value reported in Ref.~\cite{rosin:373}.

\begin{figure}[h]
\centering
\includegraphics[width=3in]{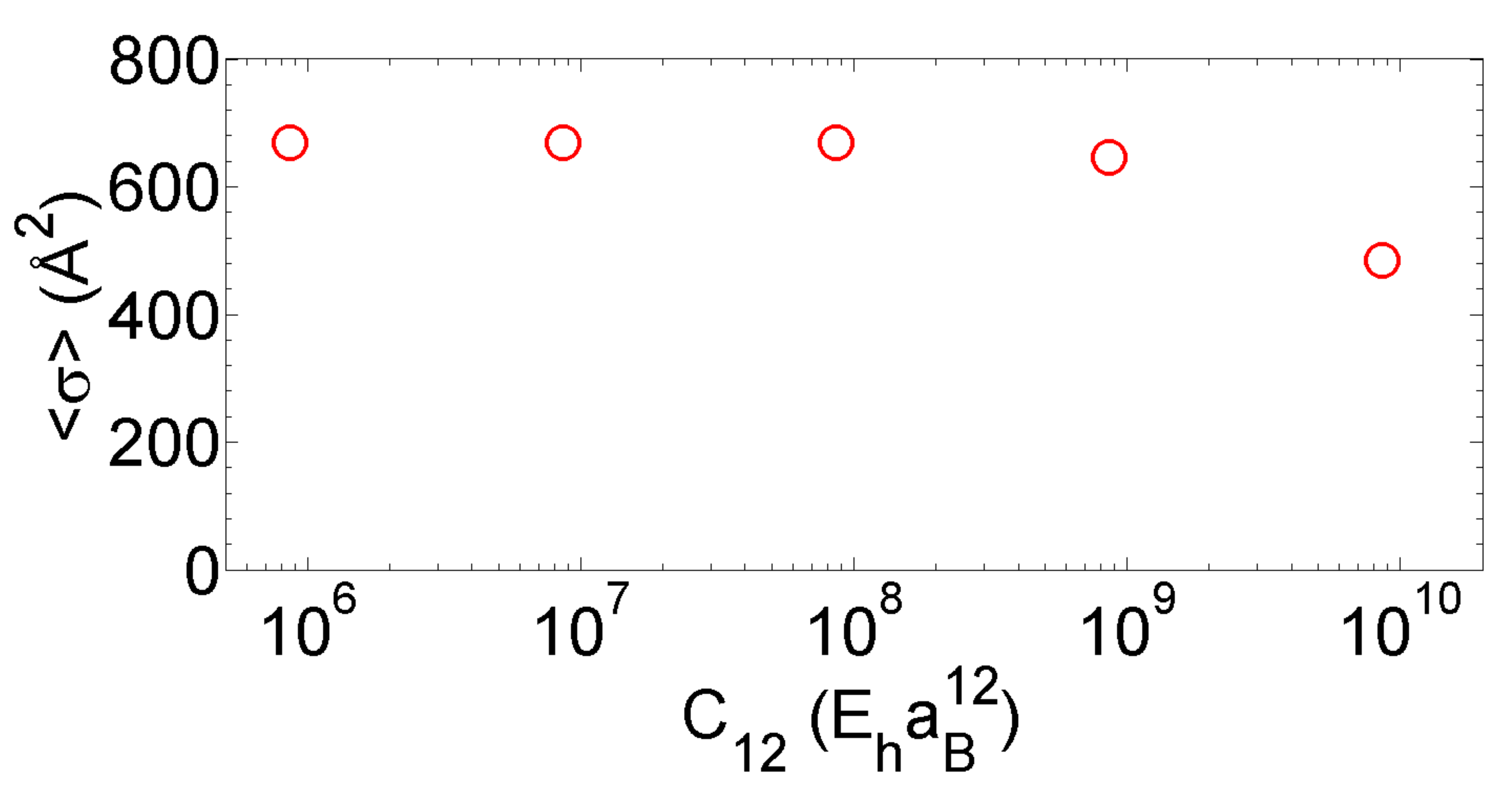}
\caption{(Color online) The theoretically computed total cross section for room temperature $\rbs$--$\arf$ collisions (averaged over a Maxwell-Boltzmann distribution at 295~K) given a Lennard-Jones inter-atomic potential as a function of the $C_{12}$ coefficient with $C_6=280~\eh \aB^6$. A value of $C_{12}=8.6 \times 10^7~\eh \aB^{12}$ results in a potential with a minimum of 50~cm$^{-1}$ below threshold.  As expected, for the collision energies characteristic of a room temperature gas, the exact value of the $C_{12}$ coefficient (at or below $10^9~\eh \aB^{12}$)  plays little role in the resulting cross section.
}
\label{fig:AvgCSVsC12}
\end{figure}

Finally, the sensitivity of the exact value of the $C_{12}$ coefficient on the computed cross section was also checked.  The results of those calculations are shown in Fig.~\ref{fig:AvgCSVsC12}.  The conclusion is clear, for any values of $C_{12}$ less than or equal to $10^{10}~\eh \aB^{12}$, more than 10 times larger than our chosen value, the cross section remains essentially unchanged (varying by less than 3\%).

\begin{figure}[h]
\centering
\includegraphics[width=3in]{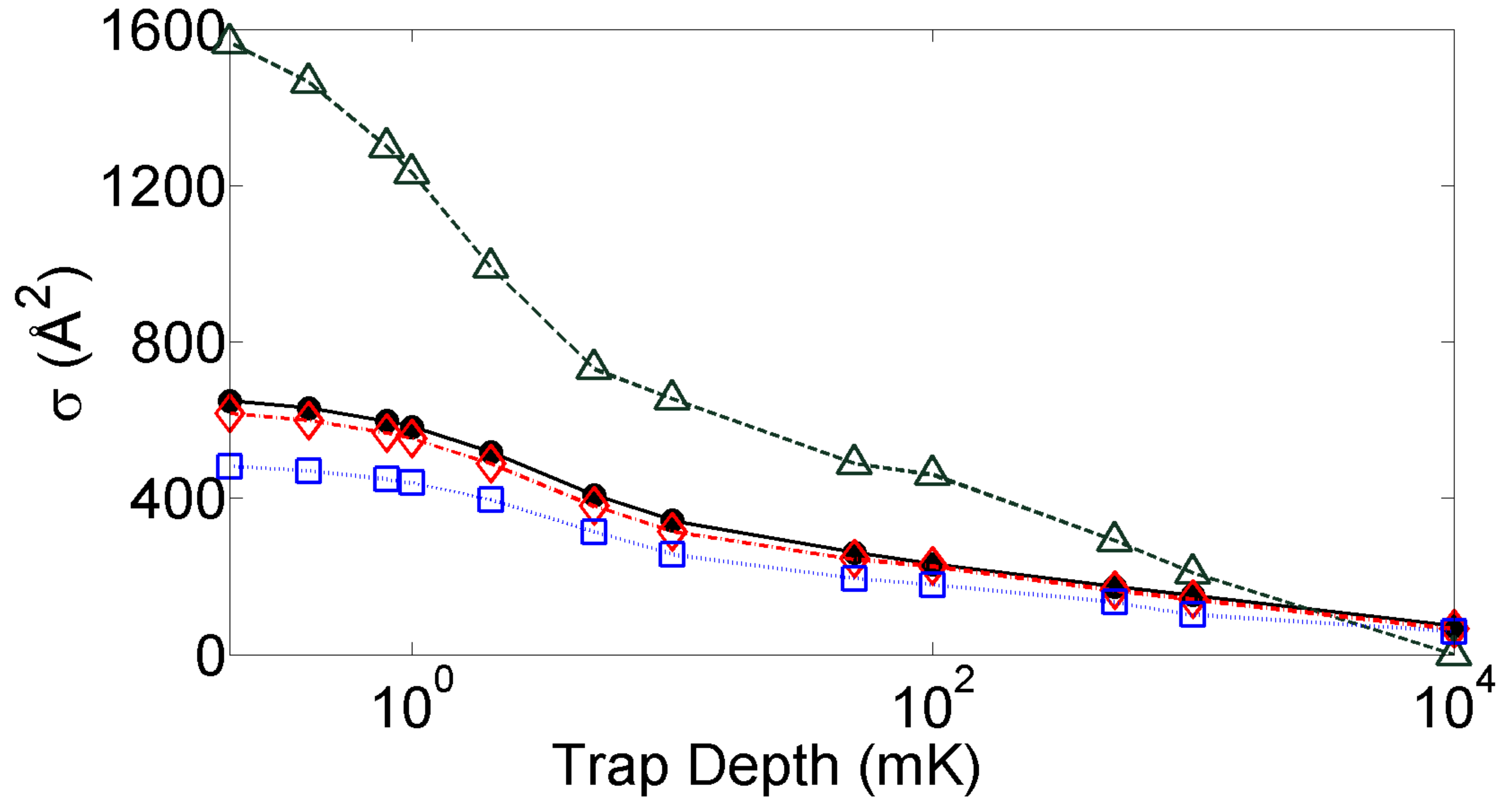}
\caption{(Color online) The theoretically computed loss cross section as a function of trap depth for $\rbs$--$\arf$ collisions of different velocities.  Loss cross sections for velocities corresponding to the average speed at 4~K (open triangles), 295~K (open diamonds), 1000~K (open squares) are shown as well as the average loss cross section for a room temperature thermal distribution (filled circles).  The variation of the loss cross section with trap depth is most pronounced for the lowest velocity collisions.  However, there still remains significant variation with trap depth for a thermal distribution at room temperature.
}
\label{fig:CSVsTrapDepth}
\end{figure}

In Fig.~\ref{fig:CSVsTrapDepth}, the theoretically computed total loss cross section of trapped $\rbs$ as a function of the trap depth for $\rbs$--$\arf$ collisions at different collision velocities is shown.  As one would expect, the variation in the loss inducing cross section with trap depth is the most pronounced at low collision velocities.  However, there still remains a significant variation in cross section for a room temperature thermal distribution of collision velocities. 

Fig.~\ref{fig:AvgCSVvsTrapDepthwithXERR} shows the theoretically computed and experimentally measured loss rate slope,  $\langle \sabarv \rangle$, for trapped $\rbs$ as a function of the trap depth.  This is the central result of this paper.  It reveals that the trap loss rate is a strongly varying function of the depth down to values below 1~mK.  In fact, the variation with depth is observed to become more pronounced at values just below 10~mK.  This behavior can be understood in the following way: approximately half of the total collision cross section arises from classical scattering (exactly half for a hard core potential) while the other half arises from quantum diffractive collisions \cite{Child}. The natural energy scale for diffractive collisions is
\be
\ed = \frac{4 \pi \hbar^2}{\ma \sigma}
\label{eq:diffenergy}
\ee
where $\ma$ is the trapped atom mass and $\sigma$ is the total collision cross section \cite{PhysRevA.60.R29}.  For Ar--Rb collisions, this energy scale is 10~mK, and a trap of this depth would exhibit roughly half the loss rate of a trap of zero depth.  The computed loss rate slope at 10~mK is 
$\langle \sabarv \rangle = 1.27 \times 10^{-9}$~cm$^3$s$^{-1}$, almost exactly half the zero trap depth loss rate slope of $2.43 \times 10^{-9}$~cm$^3$s$^{-1}$.  For trap depths below 0.01~$\ed$, the Rb loss rate differs from the zero trap depth rate by less than 1\%.  The conclusion is that by using a very shallow trap, whose depth can be easily characterized and made small compared to $\ed$, the total loss rate can be made to approach the total collision rate.  We note that the variation of the loss rates within the range of experimentally accessible trap depths provides a means to map out the differential cross section from the diffractive regime to the classical small angle regime.  Measurements of this sort will be the subject of future work.

\section{Conclusions}

In summary, the use of trapped, laser cooled rubidium atoms to measure the cross section for $\rbs$--$\arf$ collisions is investigated. Trap loss rates produced by background collisions as a function of trap depth at 1~K and below 10~mK using a magneto-optic and a quadrupole magnetic trap are studied. The retention of atoms due to a finite trap depth (i.e.~larger than the quantum diffractive collision energy scale) is found to significantly reduce the measured loss rate in comparison with the total collision rate.  Using model interaction potentials, the experimentally measured loss rates at various trap depths are found to be consistent with the known long range $C_6$ coefficient.  These results highlight the importance of minimizing the trap depth when attempting to infer the total collision cross section from measurements of atom trap loss rates.  Finally, this analysis could be used to accurately compute heating rates in magnetic traps providing a complementary approach to analytic methods \cite{cornell-1999, PhysRevA.60.R29, PhysRevA.61.033606,PhysRevA.62.063614}.

We thank Roman Krems for stimulating discussions and for his careful reading of this manuscript.
We also thank Roberto Romano for his help in fabricating the detection photodiode, low-noise amplifier. B.G. Klappauf acknowledges support from the Canadian Institute for Advanced Research. This work was supported by the Natural Sciences and Engineering Research Council of Canada (NSERC), the Canadian Foundation for Innovation (CFI), UBC, and the BCIT School of Computing and Academic Studies Professional Development Fund.


\begin{thebibliography}{33}
\expandafter\ifx\csname natexlab\endcsname\relax\def\natexlab#1{#1}\fi
\expandafter\ifx\csname bibnamefont\endcsname\relax
  \def\bibnamefont#1{#1}\fi
\expandafter\ifx\csname bibfnamefont\endcsname\relax
  \def\bibfnamefont#1{#1}\fi
\expandafter\ifx\csname citenamefont\endcsname\relax
  \def\citenamefont#1{#1}\fi
\expandafter\ifx\csname url\endcsname\relax
  \def\url#1{\texttt{#1}}\fi
\expandafter\ifx\csname urlprefix\endcsname\relax\def\urlprefix{URL }\fi
\providecommand{\bibinfo}[2]{#2}
\providecommand{\eprint}[2][]{\url{#2}}

\bibitem[{\citenamefont{Monroe et~al.}(1990)\citenamefont{Monroe, Swann,
  Robinson, and Wieman}}]{PhysRevLett.65.1571}
\bibinfo{author}{\bibfnamefont{C.}~\bibnamefont{Monroe}},
  \bibinfo{author}{\bibfnamefont{W.}~\bibnamefont{Swann}},
  \bibinfo{author}{\bibfnamefont{H.}~\bibnamefont{Robinson}}, \bibnamefont{and}
  \bibinfo{author}{\bibfnamefont{C.}~\bibnamefont{Wieman}},
  \bibinfo{journal}{Phys. Rev. Lett.} \textbf{\bibinfo{volume}{65}},
  \bibinfo{pages}{1571} (\bibinfo{year}{1990}).

\bibitem[{\citenamefont{Bjorkholm}(1988)}]{PhysRevA.38.1599}
\bibinfo{author}{\bibfnamefont{J.~E.} \bibnamefont{Bjorkholm}},
  \bibinfo{journal}{Phys. Rev. A} \textbf{\bibinfo{volume}{38}},
  \bibinfo{pages}{1599} (\bibinfo{year}{1988}).

\bibitem[{\citenamefont{Steane et~al.}(1992)\citenamefont{Steane, Chowdhury,
  and Foot}}]{Steane:92}
\bibinfo{author}{\bibfnamefont{A.~M.} \bibnamefont{Steane}},
  \bibinfo{author}{\bibfnamefont{M.}~\bibnamefont{Chowdhury}},
  \bibnamefont{and} \bibinfo{author}{\bibfnamefont{C.~J.} \bibnamefont{Foot}},
  \bibinfo{journal}{J. Opt. Soc. Am. B} \textbf{\bibinfo{volume}{9}},
  \bibinfo{pages}{2142} (\bibinfo{year}{1992}).

\bibitem[{\citenamefont{Monroe et~al.}(1993)\citenamefont{Monroe, Cornell,
  Sackett, Myatt, and Wieman}}]{PhysRevLett.70.414}
\bibinfo{author}{\bibfnamefont{C.~R.} \bibnamefont{Monroe}},
  \bibinfo{author}{\bibfnamefont{E.~A.} \bibnamefont{Cornell}},
  \bibinfo{author}{\bibfnamefont{C.~A.} \bibnamefont{Sackett}},
  \bibinfo{author}{\bibfnamefont{C.~J.} \bibnamefont{Myatt}}, \bibnamefont{and}
  \bibinfo{author}{\bibfnamefont{C.~E.} \bibnamefont{Wieman}},
  \bibinfo{journal}{Phys. Rev. Lett.} \textbf{\bibinfo{volume}{70}},
  \bibinfo{pages}{414} (\bibinfo{year}{1993}).

\bibitem[{\citenamefont{Cornell et~al.}(1999)\citenamefont{Cornell, Ensher, and
  Wieman}}]{cornell-1999}
\bibinfo{author}{\bibfnamefont{E.~A.} \bibnamefont{Cornell}},
  \bibinfo{author}{\bibfnamefont{J.~R.} \bibnamefont{Ensher}},
  \bibnamefont{and} \bibinfo{author}{\bibfnamefont{C.~E.}
  \bibnamefont{Wieman}}, \bibinfo{journal}{arXiv:cond-mat/9903109v1}
  (\bibinfo{year}{1999}).

\bibitem[{\citenamefont{Bali et~al.}(1999)\citenamefont{Bali, O\char39{}Hara,
  Gehm, Granade, and Thomas}}]{PhysRevA.60.R29}
\bibinfo{author}{\bibfnamefont{S.}~\bibnamefont{Bali}},
  \bibinfo{author}{\bibfnamefont{K.~M.} \bibnamefont{O\char39{}Hara}},
  \bibinfo{author}{\bibfnamefont{M.~E.} \bibnamefont{Gehm}},
  \bibinfo{author}{\bibfnamefont{S.~R.} \bibnamefont{Granade}},
  \bibnamefont{and} \bibinfo{author}{\bibfnamefont{J.~E.}
  \bibnamefont{Thomas}}, \bibinfo{journal}{Phys. Rev. A}
  \textbf{\bibinfo{volume}{60}}, \bibinfo{pages}{R29} (\bibinfo{year}{1999}).

\bibitem[{\citenamefont{Beijerinck}(2000{\natexlab{a}})}]{PhysRevA.61.033606}
\bibinfo{author}{\bibfnamefont{H.~C.~W.} \bibnamefont{Beijerinck}},
  \bibinfo{journal}{Phys. Rev. A} \textbf{\bibinfo{volume}{61}},
  \bibinfo{pages}{033606} (\bibinfo{year}{2000}{\natexlab{a}}).

\bibitem[{\citenamefont{Beijerinck}(2000{\natexlab{b}})}]{PhysRevA.62.063614}
\bibinfo{author}{\bibfnamefont{H.~C.~W.} \bibnamefont{Beijerinck}},
  \bibinfo{journal}{Phys. Rev. A} \textbf{\bibinfo{volume}{62}},
  \bibinfo{pages}{063614} (\bibinfo{year}{2000}{\natexlab{b}}).

\bibitem[{\citenamefont{Matherson et~al.}(2007)\citenamefont{Matherson, Glover,
  Laban, and Sang}}]{Matherson07}
\bibinfo{author}{\bibfnamefont{K.~J.} \bibnamefont{Matherson}},
  \bibinfo{author}{\bibfnamefont{R.~D.} \bibnamefont{Glover}},
  \bibinfo{author}{\bibfnamefont{D.~E.} \bibnamefont{Laban}}, \bibnamefont{and}
  \bibinfo{author}{\bibfnamefont{R.~T.} \bibnamefont{Sang}},
  \bibinfo{journal}{Rev. Sci. Instrum.} \textbf{\bibinfo{volume}{78}},
  \bibinfo{pages}{073102} (\bibinfo{year}{2007}).

\bibitem[{\citenamefont{Matherson et~al.}(2008)\citenamefont{Matherson, Glover,
  Laban, and Sang}}]{matherson08}
\bibinfo{author}{\bibfnamefont{K.~J.} \bibnamefont{Matherson}},
  \bibinfo{author}{\bibfnamefont{R.~D.} \bibnamefont{Glover}},
  \bibinfo{author}{\bibfnamefont{D.~E.} \bibnamefont{Laban}}, \bibnamefont{and}
  \bibinfo{author}{\bibfnamefont{R.~T.} \bibnamefont{Sang}},
  \bibinfo{journal}{Physical Review A (Atomic, Molecular, and Optical Physics)}
  \textbf{\bibinfo{volume}{78}}, \bibinfo{eid}{042712}
  (pages~\bibinfo{numpages}{5}) (\bibinfo{year}{2008}).

\bibitem[{\citenamefont{Dinneen et~al.}(1992)\citenamefont{Dinneen, Wallace,
  Tan, and Gould}}]{dinneen:1706}
\bibinfo{author}{\bibfnamefont{T.~P.} \bibnamefont{Dinneen}},
  \bibinfo{author}{\bibfnamefont{C.~D.} \bibnamefont{Wallace}},
  \bibinfo{author}{\bibfnamefont{K.-Y.~N.} \bibnamefont{Tan}},
  \bibnamefont{and} \bibinfo{author}{\bibfnamefont{P.~L.} \bibnamefont{Gould}},
  \bibinfo{journal}{Optics Letters} \textbf{\bibinfo{volume}{17}},
  \bibinfo{pages}{1706} (\bibinfo{year}{1992}).

\bibitem[{\citenamefont{Claessens et~al.}(2006)\citenamefont{Claessens,
  Ashmore, Sang, MacGillivray, Beijerinck, and Vredenbregt}}]{claessens:012706}
\bibinfo{author}{\bibfnamefont{B.~J.} \bibnamefont{Claessens}},
  \bibinfo{author}{\bibfnamefont{J.~P.} \bibnamefont{Ashmore}},
  \bibinfo{author}{\bibfnamefont{R.~T.} \bibnamefont{Sang}},
  \bibinfo{author}{\bibfnamefont{W.~R.} \bibnamefont{MacGillivray}},
  \bibinfo{author}{\bibfnamefont{H.~C.~W.} \bibnamefont{Beijerinck}},
  \bibnamefont{and} \bibinfo{author}{\bibfnamefont{E.~J.~D.}
  \bibnamefont{Vredenbregt}}, \bibinfo{journal}{Physical Review A (Atomic,
  Molecular, and Optical Physics)} \textbf{\bibinfo{volume}{73}},
  \bibinfo{eid}{012706} (pages~\bibinfo{numpages}{6}) (\bibinfo{year}{2006}).

\bibitem[{\citenamefont{Schappe et~al.}(1995)\citenamefont{Schappe, Feng,
  Anderson, Lin, and Walker}}]{0295-5075-29-6-002}
\bibinfo{author}{\bibfnamefont{R.~S.} \bibnamefont{Schappe}},
  \bibinfo{author}{\bibfnamefont{P.}~\bibnamefont{Feng}},
  \bibinfo{author}{\bibfnamefont{L.~W.} \bibnamefont{Anderson}},
  \bibinfo{author}{\bibfnamefont{C.~C.} \bibnamefont{Lin}}, \bibnamefont{and}
  \bibinfo{author}{\bibfnamefont{T.}~\bibnamefont{Walker}},
  \bibinfo{journal}{EPL (Europhysics Letters)} \textbf{\bibinfo{volume}{29}},
  \bibinfo{pages}{439} (\bibinfo{year}{1995}).

\bibitem[{\citenamefont{Schappe et~al.}(1996)\citenamefont{Schappe, Walker,
  Anderson, and Lin}}]{PhysRevLett.76.4328}
\bibinfo{author}{\bibfnamefont{R.~S.} \bibnamefont{Schappe}},
  \bibinfo{author}{\bibfnamefont{T.}~\bibnamefont{Walker}},
  \bibinfo{author}{\bibfnamefont{L.~W.} \bibnamefont{Anderson}},
  \bibnamefont{and} \bibinfo{author}{\bibfnamefont{C.~C.} \bibnamefont{Lin}},
  \bibinfo{journal}{Phys. Rev. Lett.} \textbf{\bibinfo{volume}{76}},
  \bibinfo{pages}{4328} (\bibinfo{year}{1996}).

\bibitem[{\citenamefont{Gensemer et~al.}(1997)\citenamefont{Gensemer,
  Sanchez-Villicana, Tan, Grove, and Gould}}]{Gensemer97}
\bibinfo{author}{\bibfnamefont{S.~D.} \bibnamefont{Gensemer}},
  \bibinfo{author}{\bibfnamefont{V.}~\bibnamefont{Sanchez-Villicana}},
  \bibinfo{author}{\bibfnamefont{K.~Y.~N.} \bibnamefont{Tan}},
  \bibinfo{author}{\bibfnamefont{T.~T.} \bibnamefont{Grove}}, \bibnamefont{and}
  \bibinfo{author}{\bibfnamefont{P.~L.} \bibnamefont{Gould}},
  \bibinfo{journal}{Physical Review A} \textbf{\bibinfo{volume}{56}},
  \bibinfo{pages}{4055} (\bibinfo{year}{1997}).

\bibitem[{\citenamefont{Kusunoki}(1967)}]{Kusunoki1967}
\bibinfo{author}{\bibfnamefont{I.}~\bibnamefont{Kusunoki}},
  \bibinfo{journal}{Bulletin of the Chemical Society of Japan}
  \textbf{\bibinfo{volume}{40}}, \bibinfo{pages}{69} (\bibinfo{year}{1967}).

\bibitem[{\citenamefont{Landau and Lifshitz}(1965)}]{Landau58}
\bibinfo{author}{\bibfnamefont{L.}~\bibnamefont{Landau}} \bibnamefont{and}
  \bibinfo{author}{\bibfnamefont{L.}~\bibnamefont{Lifshitz}},
  \emph{\bibinfo{title}{Quantum Mechanics, non-relativistic theory}}
  (\bibinfo{publisher}{Pergamon, New York}, \bibinfo{year}{1965}).

\bibitem[{\citenamefont{Helbing and Pauly}(1964)}]{Helbing64}
\bibinfo{author}{\bibfnamefont{R.}~\bibnamefont{Helbing}} \bibnamefont{and}
  \bibinfo{author}{\bibfnamefont{H.}~\bibnamefont{Pauly}},
  \bibinfo{journal}{Zeitschrift fuer Physik} \textbf{\bibinfo{volume}{179}},
  \bibinfo{pages}{16} (\bibinfo{year}{1964}).

\bibitem[{\citenamefont{Anderson}(1974)}]{anderson:2680}
\bibinfo{author}{\bibfnamefont{R.~W.} \bibnamefont{Anderson}},
  \bibinfo{journal}{The Journal of Chemical Physics}
  \textbf{\bibinfo{volume}{60}}, \bibinfo{pages}{2680} (\bibinfo{year}{1974}).

\bibitem[{\citenamefont{Fagnan}(2009)}]{fagnan}
\bibinfo{author}{\bibfnamefont{D.}~\bibnamefont{Fagnan}},
  \bibinfo{type}{Honour's Thesis}, \bibinfo{institution}{University of British
  Columbia} (\bibinfo{year}{2009}).

\bibitem[{\citenamefont{Child}(1974)}]{Child}
\bibinfo{author}{\bibfnamefont{M.}~\bibnamefont{Child}},
  \emph{\bibinfo{title}{Molecular Collision Theory}}
  (\bibinfo{publisher}{Academic, New York}, \bibinfo{year}{1974}).

\bibitem[{\citenamefont{Markovi\'c}(2003)}]{Markovic}
\bibinfo{author}{\bibfnamefont{N.}~\bibnamefont{Markovi\'c}},
  \bibinfo{type}{unpublished}, \bibinfo{institution}{Chalmers University of
  Technology, G\"oteborg, Sweden} (\bibinfo{year}{2003}).

\bibitem[{\citenamefont{Johnson}(1973)}]{Johnson:445}
\bibinfo{author}{\bibfnamefont{B.}~\bibnamefont{Johnson}},
  \bibinfo{journal}{Journal of Computational Physics}
  \textbf{\bibinfo{volume}{13}}, \bibinfo{pages}{445} (\bibinfo{year}{1973}).

\bibitem[{\citenamefont{Ladouceur et~al.}(2009)\citenamefont{Ladouceur,
  Klappauf, Dongen, Rauhut, Schuster, Mills, Jones, and
  Madison}}]{Ladouceur:09}
\bibinfo{author}{\bibfnamefont{K.}~\bibnamefont{Ladouceur}},
  \bibinfo{author}{\bibfnamefont{B.~G.} \bibnamefont{Klappauf}},
  \bibinfo{author}{\bibfnamefont{J.~V.} \bibnamefont{Dongen}},
  \bibinfo{author}{\bibfnamefont{N.}~\bibnamefont{Rauhut}},
  \bibinfo{author}{\bibfnamefont{B.}~\bibnamefont{Schuster}},
  \bibinfo{author}{\bibfnamefont{A.~K.} \bibnamefont{Mills}},
  \bibinfo{author}{\bibfnamefont{D.~J.} \bibnamefont{Jones}}, \bibnamefont{and}
  \bibinfo{author}{\bibfnamefont{K.~W.} \bibnamefont{Madison}},
  \bibinfo{journal}{J. Opt. Soc. Am. B} \textbf{\bibinfo{volume}{26}},
  \bibinfo{pages}{210} (\bibinfo{year}{2009}).

\bibitem[{\citenamefont{Steck}()}]{Steck08}
\bibinfo{author}{\bibfnamefont{D.~A.} \bibnamefont{Steck}},
  \bibinfo{note}{rubidium 87 D Line Data, available online at
  http://steck.us/alkalidata (revision 2.0.1, 2 May 2008).}

\bibitem[{\citenamefont{Hoffmann et~al.}(1996)\citenamefont{Hoffmann, Bali, and
  Walker}}]{PhysRevA.54.R1030}
\bibinfo{author}{\bibfnamefont{D.}~\bibnamefont{Hoffmann}},
  \bibinfo{author}{\bibfnamefont{S.}~\bibnamefont{Bali}}, \bibnamefont{and}
  \bibinfo{author}{\bibfnamefont{T.}~\bibnamefont{Walker}},
  \bibinfo{journal}{Phys. Rev. A} \textbf{\bibinfo{volume}{54}},
  \bibinfo{pages}{R1030} (\bibinfo{year}{1996}).

\bibitem[{\citenamefont{Bagnato et~al.}(2000)\citenamefont{Bagnato, Marcassa,
  Miranda, Muniz, and de~Oliveira}}]{PhysRevA.62.013404}
\bibinfo{author}{\bibfnamefont{V.~S.} \bibnamefont{Bagnato}},
  \bibinfo{author}{\bibfnamefont{L.~G.} \bibnamefont{Marcassa}},
  \bibinfo{author}{\bibfnamefont{S.~G.} \bibnamefont{Miranda}},
  \bibinfo{author}{\bibfnamefont{S.~R.} \bibnamefont{Muniz}}, \bibnamefont{and}
  \bibinfo{author}{\bibfnamefont{A.~L.} \bibnamefont{de~Oliveira}},
  \bibinfo{journal}{Phys. Rev. A} \textbf{\bibinfo{volume}{62}},
  \bibinfo{pages}{013404} (\bibinfo{year}{2000}).

\bibitem[{\citenamefont{Petrich et~al.}(1995)\citenamefont{Petrich, Anderson,
  Ensher, and Cornell}}]{PhysRevLett.74.3352}
\bibinfo{author}{\bibfnamefont{W.}~\bibnamefont{Petrich}},
  \bibinfo{author}{\bibfnamefont{M.~H.} \bibnamefont{Anderson}},
  \bibinfo{author}{\bibfnamefont{J.~R.} \bibnamefont{Ensher}},
  \bibnamefont{and} \bibinfo{author}{\bibfnamefont{E.~A.}
  \bibnamefont{Cornell}}, \bibinfo{journal}{Phys. Rev. Lett.}
  \textbf{\bibinfo{volume}{74}}, \bibinfo{pages}{3352} (\bibinfo{year}{1995}).

\bibitem[{\citenamefont{Weiner et~al.}(1999)\citenamefont{Weiner, Bagnato,
  Zilio, and Julienne}}]{Weiner99}
\bibinfo{author}{\bibfnamefont{J.}~\bibnamefont{Weiner}},
  \bibinfo{author}{\bibfnamefont{V.~S.} \bibnamefont{Bagnato}},
  \bibinfo{author}{\bibfnamefont{S.}~\bibnamefont{Zilio}}, \bibnamefont{and}
  \bibinfo{author}{\bibfnamefont{P.~S.} \bibnamefont{Julienne}},
  \bibinfo{journal}{Rev. Mod. Phys.} \textbf{\bibinfo{volume}{71}},
  \bibinfo{pages}{1} (\bibinfo{year}{1999}).

\bibitem[{\citenamefont{Rosin and Rabi}(1935)}]{rosin:373}
\bibinfo{author}{\bibfnamefont{S.}~\bibnamefont{Rosin}} \bibnamefont{and}
  \bibinfo{author}{\bibfnamefont{I.}~\bibnamefont{Rabi}},
  \bibinfo{journal}{Phys. Rev.} \textbf{\bibinfo{volume}{48}},
  \bibinfo{pages}{373} (\bibinfo{year}{1935}).

\bibitem[{\citenamefont{Dalgarno and Davison}(1967)}]{dalgarno:479}
\bibinfo{author}{\bibfnamefont{A.}~\bibnamefont{Dalgarno}} \bibnamefont{and}
  \bibinfo{author}{\bibfnamefont{W.}~\bibnamefont{Davison}},
  \bibinfo{journal}{Molecular Physics} \textbf{\bibinfo{volume}{13}},
  \bibinfo{pages}{479} (\bibinfo{year}{1967}).

\bibitem[{\citenamefont{Mahan}(1968)}]{mahan:950}
\bibinfo{author}{\bibfnamefont{G.}~\bibnamefont{Mahan}}, \bibinfo{journal}{The
  Journal of Chemical Physics} \textbf{\bibinfo{volume}{48}},
  \bibinfo{pages}{950} (\bibinfo{year}{1968}).

\bibitem[{\citenamefont{Krems}(2009)}]{romanexplain}
\bibinfo{author}{\bibfnamefont{R.}~\bibnamefont{Krems}} (\bibinfo{year}{2009}),
  \bibinfo{note}{private communication}.

\end{thebibliography}

\end{document}